# Improved thermal stability of dielectric properties and energy storage properties of lead-free relaxor $Ba_{(1-x)}La_xTi_{0.89}Sn_{0.11}O_3$ ceramics.

S. Khardazi[1*], D. Mezzane[1,2], M. Amjoud[1], N. Novak[3], E. Choukri[1], S. Lyubchyk[4], Z. Kutnjak[3], B. Rožič[3], S. Terenchuk[5] and I. Lukyanchuk[5]

[1] IMED-Lab, Cadi-Ayyad University, Faculty of Sciences and Technology, Department of Applied Physics, Marrakech, 40000, Morocco

[2] Laboratory of Physics of Condensed Matter (LPMC), University of Picardie Jules Verne, Scientific Pole, 33 Rue Saint-Leu, Amiens, Cedex 1 80039, France

[3] Jožef Stefan Institute, Ljubljana, 1000, Slovenia

[4] DeepTechLab, Faculty of Engineering, Lusofona University, 376 Campo Grande, 1749-024, Lisbon, Portugal

[5] Kyiv National University of Construction and Architecture, 31, Povitroflotsky Ave., 03680, Kyiv, Ukraine

Abstract

Lead-free perovskite ceramics $Ba_{(1-x)}La_xTi_{0.89}Sn_{0.11}O_3$ with x= 0, 0.015, 0.025, and 0.035 (BLTSnx) were synthesized by solid-state reaction method. The enhanced energy storage properties were studied across the ferroelectric relaxor conversion. The pure perovskite structure of all prepared samples was confirmed by X-ray diffraction (XRD) analysis and the Raman spectroscopy technique. Scanning electron microscopy micrographs show a drastic decrease in the grain size in La-doped ceramics. Furthermore, the doped samples exhibit smeared phase transition and frequency dispersion in dielectric permittivity, typical for relaxors. The P-E hysteresis loops become thin with the increase of the La-doping, demonstrating the ferroelectric-relaxor phase conversion. The enhanced recovered energy density and the energy storage efficiency of 158.8 mJ/cm$^3$ and 82.73 % were observed in the BLTSn2.5 sample at room temperature, respectively. These properties indicate that environmentally friendly BLTSn2.5 is a viable candidate for energy-storage capacitor applications near room temperature.

* Corresponding author.

E-mail address: khardazzisaid@gmail.com, said.khardazi@ced.uca.ma  (S. Khardazi).

Tel: +212 608219671

# 1. Introduction

Nowadays, dielectric ceramics have attracted significant attention among various energy storage materials due to their high power density, wide temperature range stability, and fast charge/discharge capabilities, which make them an essential component of energy storage devices that are essential in power systems and electronic equipment [1-3]. Current studies concentrate on developing innovative ferroelectric materials with high electrocaloric, piezoelectric, and energy-storage properties near room temperature. Traditionally, lead-based ceramics, such as lead zirconate titanate (PZT) and magnesium niobate lead titanate (PMN-PT) have been widely studied for energy storage applications due to their superior dielectric properties [4]. However, the environmental and health hazards associated with lead have driven the scientific community to explore lead-free alternatives.

Developing lead-free ceramics is crucial to creating environmentally friendly and sustainable energy storage systems. As is known, lead-free ceramics such as barium titanate ($BaTiO_3$), bismuth sodium titanate ($Bi_{0.5}Na_{0.5}TiO_3$), and potassium sodium niobate ($K_{0.5}Na_{0.5}NbO_3$), have shown promise as ecological replacements for lead-based materials [5, 6]. These materials exhibit high dielectric permittivity and good ferroelectric properties, making them suitable candidates for various applications, including capacitors, actuators, and transducers [7]. Theoretically, maximum polarization ($P_{max}$), breakdown strength ($E_b$), and remanent polarization ($P_r$) of dielectric materials affect the recovered energy storage density ($W_{rec}$) and efficiency (η) of dielectric capacitors [8]. It is commonly recognized that optimizing the energy storage capability of capacitors requires achieving low $P_r$, high $P_{max}$, and $E_b$ simultaneously [9], [10]. Unfortunately, these ceramics have two major disadvantages that limit applications in energy storage: low breakdown field strength and significant hysteresis loss [11]. Consequently, doping with other metal oxides or creating solid solutions with different compounds can diminish their ferroelectric properties while improving their relaxor ferroelectricity [12]. However, achieving simultaneously high η and a large $P_{max}$ is still challenging [13]. High η relates to the polar nanoregions (PNRs) in RFE materials, but large $P_{max}$ is generally associated with macro domains in ferroelectric materials [9].

The lead-free BT-based $BaTi_{0.89}Sn_{0.11}O_3$ exhibits a high dielectric constant, a superior piezoelectric coefficient, a significant electrocaloric effect, and improved energy storage capabilities [14, 15]. Therefore, the notable dielectric, energy-storage, and electrocaloric characteristics are apparent only within a limited temperature range. To overcome this drawback, researchers have concentrated on systems with a diffuse transition typically found in ferroelectric relaxors that expand the operational temperature range [16]. It has been reported that substituting $La^{3+}$ for $Ba^{2+}$ leads to the creation of B-site vacancies that compensate for the imbalanced charges. Thus, by controlling the microstructure, relaxor characteristics can be adjusted by using both homovalent and heterovalent replacements at the A site (La) and B site (Sn) [17, 18]. Furthermore, Kumar *et al.* [19] have reported an enhancement of the energy storage density of 0.492 $J/cm^3$ along with a relatively low η of ~63%, in La-doped $BaTi_{0.95}Sn_{0.05}O_3$ ceramics. Furthermore, a high energy storage density of 0.91 $J/cm^3$ and η of 69% in $[(Bi_{0.5}Na_{0.5})_{0.93}Ba_{0.07}]_{1-x}La_xTiO_3$ ceramics at x = 0.04 was registered by Xu *et al.* [20]. Recently, several studies have confirmed that lead-free-based relaxor ceramics have improved energy storage properties [20- 22].

In this paper, the $La^{3+}$ doped $BaTi_{0.89}Sn_{0.11}O_3$ ceramics were prepared using solid-state reactions. The effect of La substitution on phase transition, dielectric relaxor behavior, and energy storage property was studied.

## 2. Experimental details

$Ba_{(1-x)}La_xTi_{0.89}Sn_{0.11}O_3$ ceramics ( abbreviated as $B_{(1-x)}L_xTSn$, x = 0, 1.5, 2.5, and 3.5 mol%) were synthesized using the conventional solid-state reaction method. High-purity chemicals such as $BaCO_3$, $La_2O_3$, $TiO_2$, and $SnO_2$ were chosen as starting materials. They were weighed stoichiometrically and mixed with ethanol in an agate mortar for 2h. Then, the mixtures were calcined at 1150 °C for 13h in air. Next, the calcined powders were mixed with 5 wt.% polyvinyl alcohol (PVA) and compacted uniaxially to form cylindrical pellets with a diameter of approximately 12 mm. All the green-pressings were then sintered at 1350°C for 6h at the heating rate of 5 °C/min.

The X-ray diffractometer PANalytical X-Pert Pro under Cu-Kα radiation with λ ~ 1.540598 Å at room temperature was used to determine the crystalline phases of ceramics. The microstructure of the sintered pellets was investigated by using the TESCAN VEGA3 Scanning Electron Microscope (SEM). Raman spectra for all the ceramics were recorded using a micro-Raman RENISHAW with a CCD detector and green laser excitation of 532 nm. The dielectric properties were tested by the Solartron Impedance Analyzer SI-1260 in the frequency range 100 Hz – 1 MHz and temperature interval from -90 °C to 250 °C.

Temperature-dependent P–E hysteresis curves were measured at 200 Hz using a ferroelectric test system (PolyK Technologies State College, PA, USA).

## 3. Results and discussions

### 3.1 Structural and microstructural properties

XRD and Raman methods at room temperature were used to study the phase structure of $Ba_{(1-x)}La_xTi_{0.89}Sn_{0.11}O_3$ ceramics. Fig.1a illustrates the XRD patterns of $Ba_{(1-x)}La_xTi_{0.89}Sn_{0.11}O_3$ (x = 0, 1.5, 2.5 and 3.5) ceramics at room temperature. All samples exhibit one typical perovskite structure without any second phase, suggesting that $La^{3+}$ has been incorporated into the lattice to form the solid solution. The reflection peaks corresponded to the standard perovskite structure of $BaTiO_3$ and the vertical lines indicate the standard diffraction peaks of $BaTiO_3$ with a tetragonal phase (T, space group: *P4mm*) *(PDF#00-005-0626)*. In pure BTSn, the tetragonal signature is represented by the splitting of the $(002)_T/(200)_T$ peak at $2\Theta = 45°$ with a ratio less than 1. The observed peak transforms into a single peak and becomes wider for La-doped samples, indicating a shrank lattice and phase transformation in the BLTSn samples. The merging of the (002) and (200) peaks in the $2\theta$ range of 44–46° indicates the dominant pseudo-cubic phase in the BLTSn ceramics at room temperature [23, 24]. Fig.1(b) displays the enlarged reflection at the peak position ($2\Theta = 45°$) for all compositions, demonstrating the right shifting of peaks with La-doping for BLTSn 1.5 and BLTSn 2.5. The diffraction peak shifting is due to the substitution of higher ionic radius $Ba^{2+}$ (1.61 Å) compared to $La^{3+}$ (1.32 Å). The (200) peak shifts to a larger angle due to the replacement of the larger ions with those with smaller radii due to reduced interplanar spacing [25]. The shifting of the (200) peak to a lower angle in BLTSn 3.5 demonstrated an increase in lattice parameters, which could be due to the substitution in the B site of lower ionic radius $Ti^{4+}$ (0.605 Å) as compared to $La^{3+}$ (1.32 Å).

To get insight into the microstrain generated due to the substitution of $Ba^{2+}$ by $La^{3+}$, the Williamson-Hall (W–H) method was employed using the following formula [26]:

$$\beta_T cos\theta = \frac{K\lambda}{D} + 4\varepsilon\, sin\theta, \qquad (1)$$

The parameters K, D, θ, λ, and ε indicate the particle shape factor (~0.94), effective crystallite size, diffraction angle, the wavelength of the X-ray source, and the effective strain generated in the samples, respectively. Fig.2 shows the plots of $\beta_T\cos\theta$ vs. $\sin\theta$ for $Ba_{(1-x)}La_xTi_{0.89}Sn_{0.11}O_3$ ceramics. Based on the W-H method, the microstrain values obtained are $1.17\times10^{-3}$, $2.54\times10^{-3}$, $2.78\times10^{-3}$ and $1.38\times10^{-3}$ for BTSn, BLTSn 1.5, BLTSn 2.5 and BLTSn 3.5, respectively. It is noticeable that the microstrain in all the doped samples is relatively higher than that of the pure BTSn.

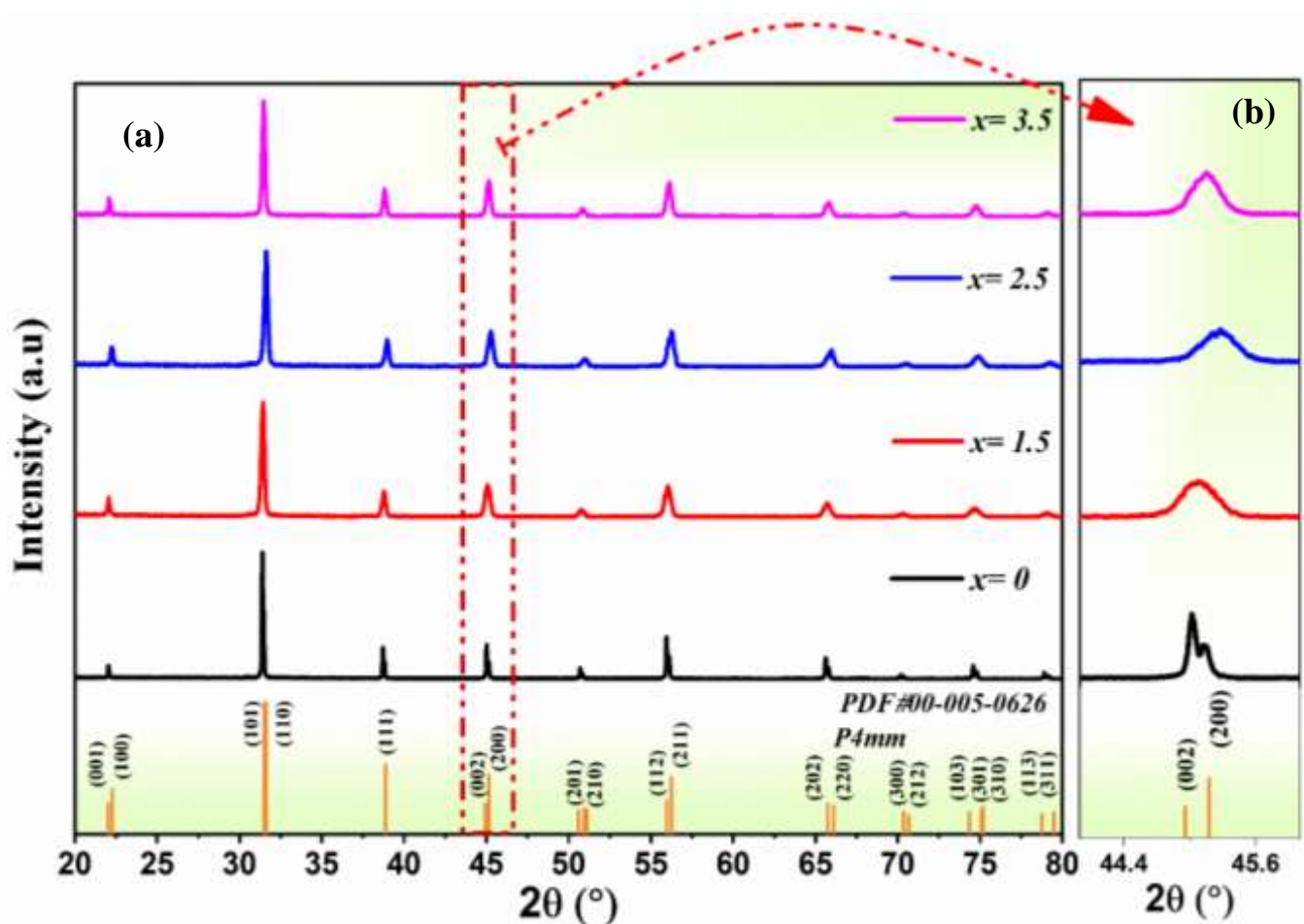

Figure 1: (a) The XRD patterns of the $Ba_{(1-x)}La_xTi_{0.89}Sn_{0.11}O_3$ ceramics, (b) the magnified (200) diffraction peaks.

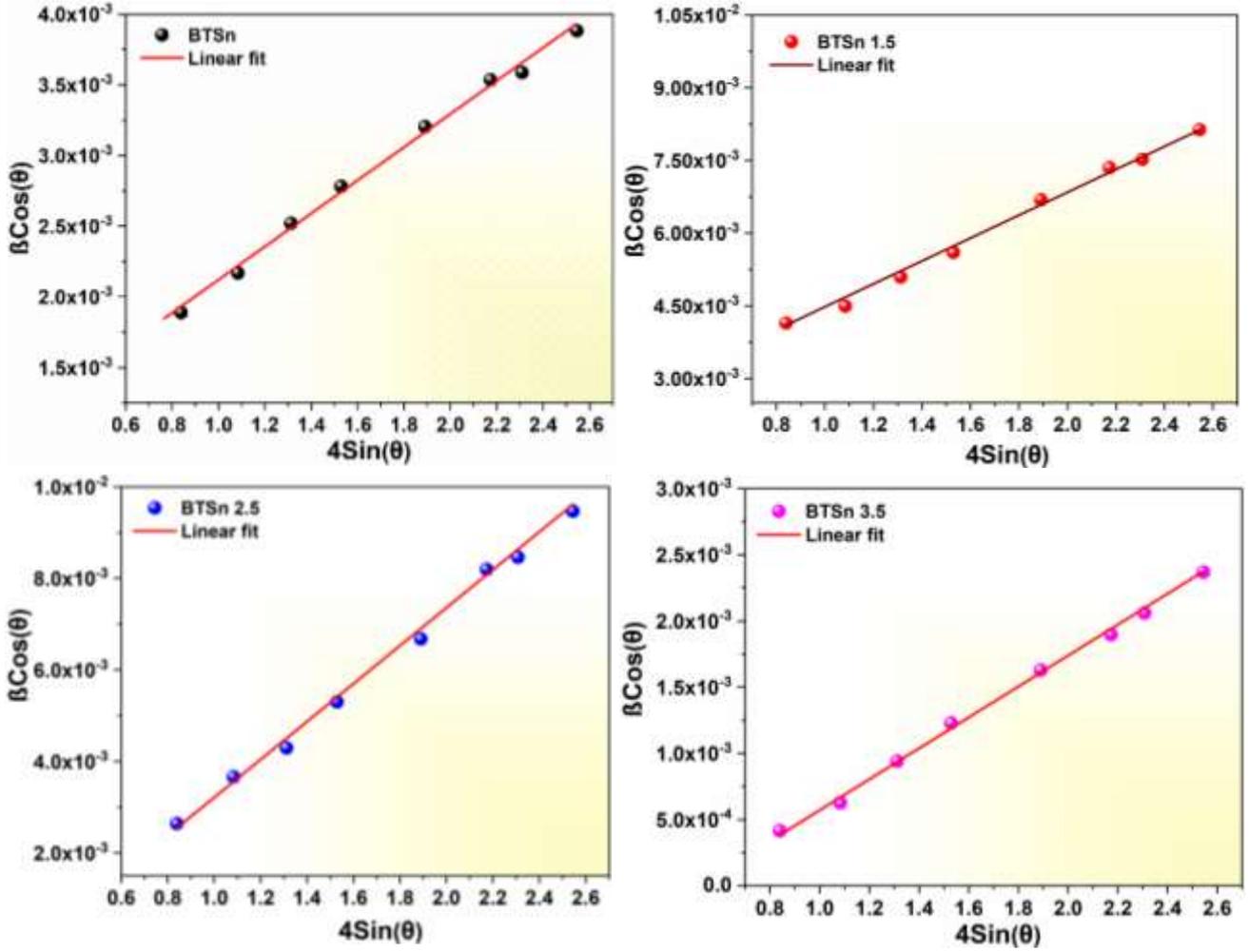

Figure 2: Williamson-Hall plots of $Ba_{(1-x)}La_xTi_{0.89}Sn_{0.11}O_3$ ceramics.

Raman spectroscopy is a versatile and powerful tool for studying ferroelectric ceramics. It provides detailed information about the vibrational modes, phase transitions, and domain structures. Figure-3a represents room temperature Raman spectra of $Ba_{(1-x)}La_xTi_{0.89}Sn_{0.11}O_3$ ceramics recorded over the 70–1000 cm$^{-1}$ wavenumber range. As shown in Fig.3a, these Raman active modes are common in the BT Raman spectra, as reported in some previous studies. The pure BTSn sample's Raman spectrum displays the following dominating modes centers at 116, 187, 260, 309, 519, and 721 cm$^{-1}$, with one anti-resonance dip at 177 cm$^{-1}$. The band detected at around 309 cm$^{-1}$ is a distinctive characteristic of the tetragonal phase in ferroelectric perovskite oxides and is attributed to a combined mode [$B_1$, E (TO + LO)]. The band situated at 721 cm$^{-1}$ is attributed to the low-intensity $A_1$/E (LO) phonon mode. Furthermore, an interference effect at 177 cm$^{-1}$ appears only in the tetragonal phase [27, 28]. With an increase in La-doping in BTS ceramics, the sharp resonance dip at 177 cm$^{-1}$ decreases. In addition, the intensity of the sharp band at 309 cm$^{-1}$ starts to decrease, especially at 3.5 atomic % of La-doping in BTSn [29]. This demonstrates that the cubic and tetragonal phases coexist at room temperature and that the volume fraction of cubic phases rises as La doping increases.

The most noticeable change in this composition, resulting from La substitution at the Ba site in the BTSn ceramic sample, is the appearance of an extra peak at 837 cm$^{-1}$. The peak at 837 cm$^{-1}$ corresponds to the $A_{1g}$ to the asymmetric breathing mode of the TiO$_6$ octahedra characteristic of the polar nano regions [30]. The same mode was observed by Feteira *et al* [31] at 823 cm$^{-1}$ in the simultaneous incorporation of La$^{3+}$ and Y$^{3+}$ ions in (1-x)BaTiO$_3$ -xLaYO$_3$ ceramics. Figure-3b shows the variation of the FWHM at a wavenumber of 519 cm$^{-1}$. With increasing the La content into the BTSn lattice, the Raman mode broadens due to structural disorder caused by substituting of La ions, which have different valence states and ionic radii. One can see that BLTSn 2.5 exhibits the highest FWHM, possibly due to the high lattice strain induced by the small grain size and the multi-phase coexistence. This result could be due to the low tetragonality shown by the lowest intensity of the Raman peaks at BLTSn 2.5, indicated by the highest value of the FWHM [32].

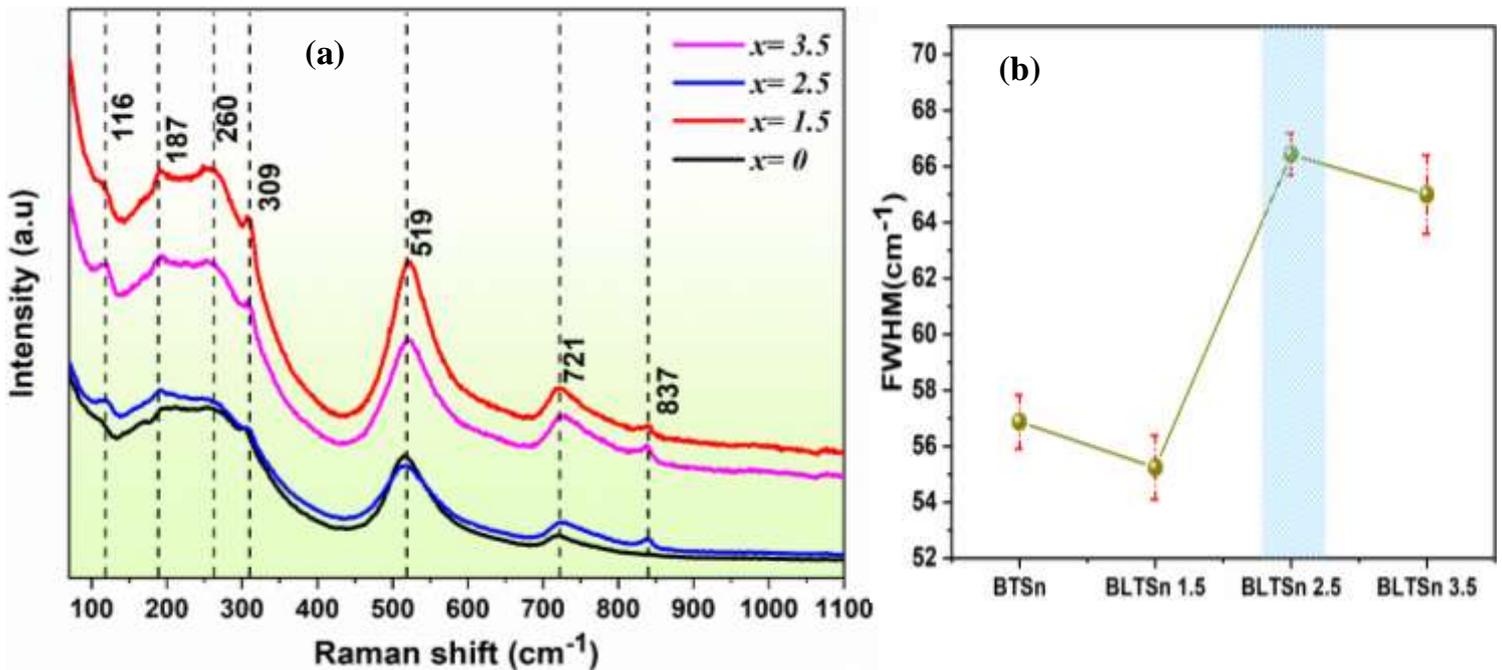

Figure 3: (a) Raman spectra, (b) evolution of WWHM at wavenumber 519 cm$^{-1}$ of Ba$_{(1-x)}$La$_x$Ti$_{0.89}$Sn$_{0.11}$O$_3$ samples at room temperature.

Fig.4 displays the SEM pictures of Ba$_{(1-x)}$La$_x$Ti$_{0.89}$Sn$_{0.11}$O$_3$ ceramics. All the samples showed well-grown grains, with no abnormal grain growth. One can see that the pure BTSn sample exhibits a larger grain size of 30 µm. The increased mass transport phenomenon caused by the creation of oxygen vacancies may be the source of the higher grain size in the pure BTSn sample. It is known that high-temperature sintering (≥1350°C) produces these oxygen vacancies in ceramics [33]. From SEM micrographs, the grain size decreased sharply with increased La content. This indicates that a donor dopant might inhibit grain growth and provide a more uniform distribution of grain sizes in the resulting solid solution. It is important to note that all doped ceramics displayed a small grain size in the ~ 0.2-1µm range. According to published reports, incorporating La ions at the Ba-site reduces the overall diffusion rate during the sintering process and may suppress the oxygen vacancies [34, 35]. In addition, the BLTSn 1.5 sample shows pores that decrease as La substitution increases and a slight increase in the grain size of the BLTSn 3.5 sample. The same observation was found by Gao.*et al* [35] in $(Ba_{1-1.5x} La_x)(Ti_{0.99}Mn_{0.01})O_3$ system and Kumar.*et al* [30] in La-doped $Ba(Ti_{0.8}Zr_{0.2})O_3$ ceramics. Furthermore, decreasing the grain size improves the density, which raises the grain boundary density, effectively impeding the migration of charge within the ceramic [30]. Therefore, this development helps to improve the energy storage performances of BLTSn samples.

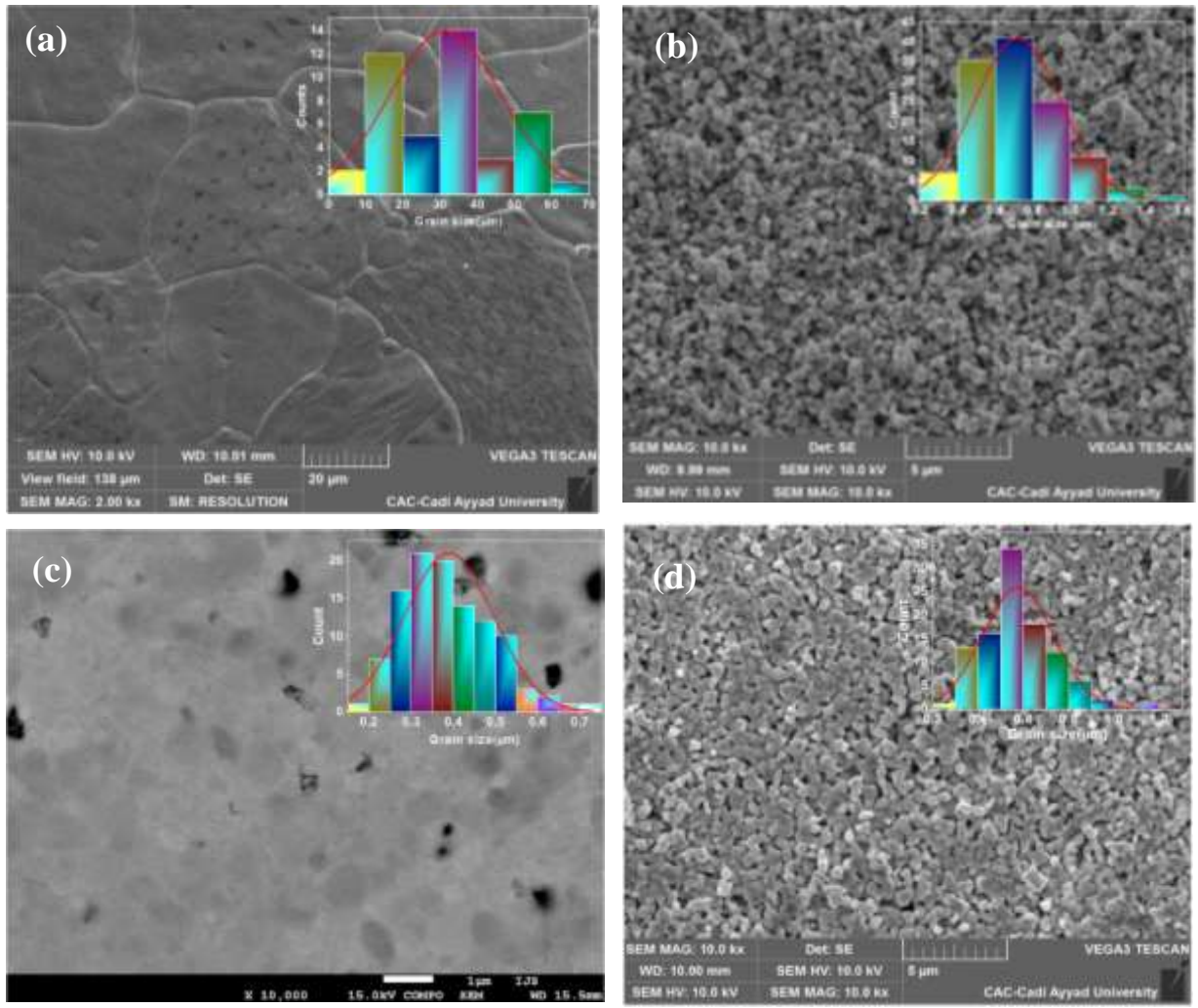

Figure 4: SEM micrographs of $Ba_{(1-x)}La_xTi_{0.89}Sn_{0.11}O_3$ ceramics. The inset of each SEM picture shows the corresponding grain size distribution.

*3.2 Dielectric properties*

Fig. 5 shows the temperature dependence of the dielectric constant ($\varepsilon$) and the dielectric loss (tgδ) for all the investigated compositions at different frequencies to examine the phase transition in the four samples under study. All the examined samples exhibit dielectric anomaly describing the Tetragonal–Cubic (Curie temperature ($T_C$)) phase transition, which indicates the Ferroelectric–Paraelectric (FE–PE) structural change. These results corroborate those obtained by XRD and Raman spectroscopy analysis. One can see that the dielectric constant decreases significantly with increasing the La content, most likely due to the smaller grain size in the doped samples as shown in the SEM micrographs. As shown in Fig.5, the phase transition at Curie temperature broadens with increasing the La concentration. For the pure BTSn sample, the frequency dispersion around dielectric maxima is weak, suggesting the presence of the macroscopic ferroelectric domains. In contrast, the La-doped samples exhibit slightly suppressed and shifted dielectric maxima ($\varepsilon_m$) to lower temperatures and increased frequency dispersion. The significant dielectric relaxor behavior of La-doped samples could be due to the increase of system disorder created by the substitution of Ba by La, leading to the creation of polar nano-regions (PNRs) [36, 37]. As the amount of La increases, the phase transitions from the tetragonal to the pseudocubic phase, creating relaxor ferroelectrics. The creation of these nanodomains (PNRs) could reduce the $\varepsilon_m$ in La-doped samples. This result could be explained by the fact that the long-range order is destroyed when ions with different ionic valence states and radii recombine at sites A or B [38, 39]. These findings suggest that the La substitution improves the degree of diffuseness of the phase transition and ferroelectric relaxor behavior.

On the other hand, the dielectric losses decrease with increasing the La content and remain below 0.02 for 1kHz. The small dielectric losses are beneficial to enhancing the energy storage efficiency (η) [40]. Furthermore, above 100 °C the tanδ increases with temperature, mainly due to generating more oxygen vacancies and defects in $Ba_{(1-x)}La_xTi_{0.89}Sn_{0.11}O_3$ ceramics at high temperatures.

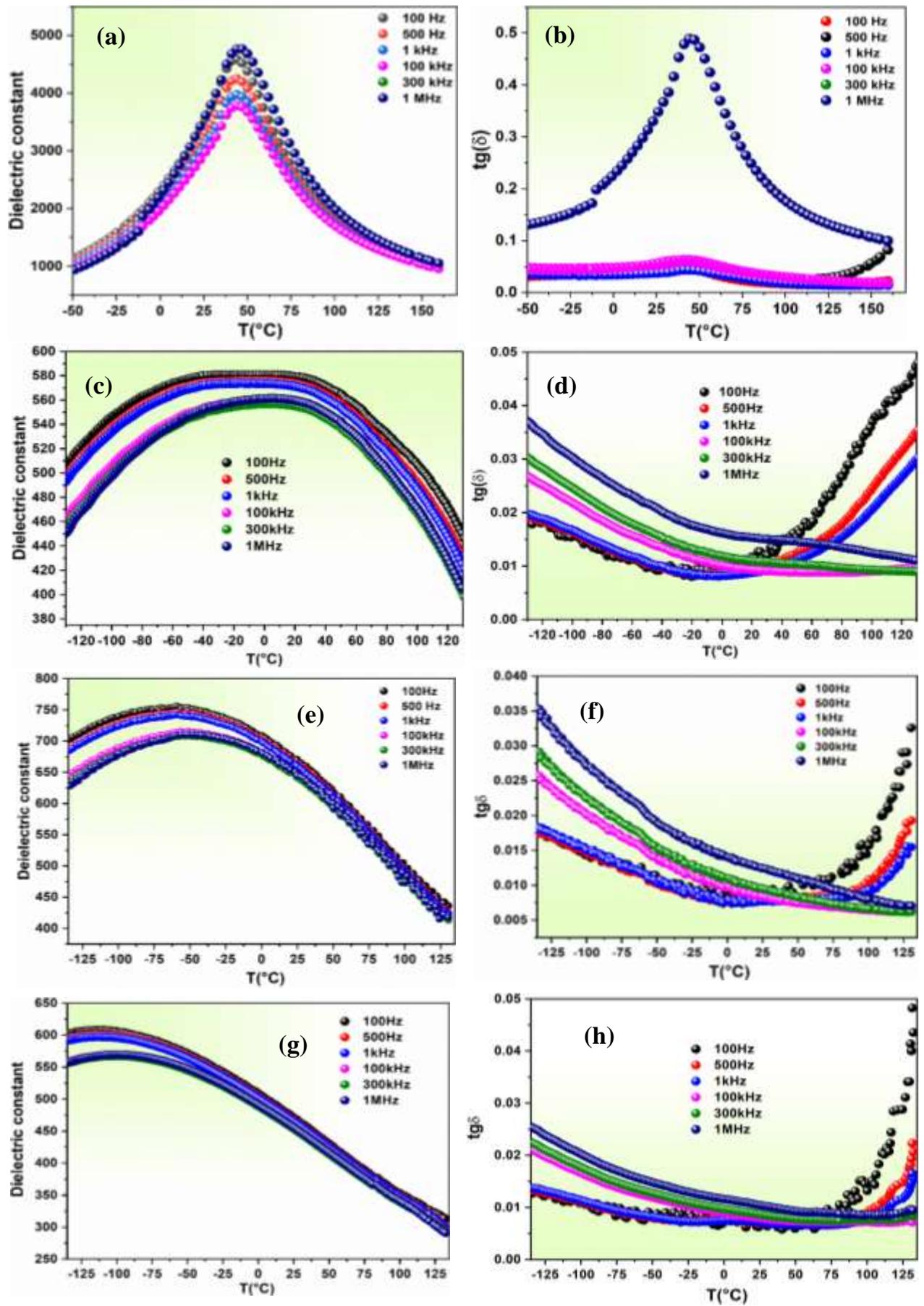

Figure 5: Temperature dependence of dielectric constant and dielectric loss at various

For further information about the phase transition, Fig.6 illustrates the relationship between temperature and inverse dielectric constant for $Ba_{(1-x)}La_xTi_{0.89}Sn_{0.11}O_3$ samples at 1 kHz frequency. In the paraelectric region, the thermal variation of the dielectric constant could be explained by Curie-Weiss law :

$$\frac{1}{\varepsilon} = \frac{(T-T_0)}{C}, \qquad (2)$$

Where respectively $T_0$ and C are the Curie-Weiss temperature and Curie-Weiss constant, and $\varepsilon$ is the real part of the dielectric constant. One can see that, the experimental data are well-fitted using Eq. (2) and the Curie constant of all samples is on the order of $10^5$ K, suggesting that at high temperatures in the paraelectric phase, the displacive phase transition is driven [41]. Furthermore, it is possible to evaluate the relaxor behavior of the diffuse phase transition (above $T_C$) using the modified Curie-Weiss relationship. This law was suggested by Uchino and Nomura and given by Eq.(3) as an updated version of the Curie-Weiss equation [42]:

$$\frac{1}{\varepsilon} - \frac{1}{\varepsilon_m} = \frac{(T-T_m)^\gamma}{C}, \quad T > T_m \qquad (3)$$

Among them, $\varepsilon_m$ represents the maximum dielectric constant at the transition temperature $T_m$, C and γ (degree of the transition diffuseness) are constants. For γ = 1, Eq. 3 describes a completely non-diffusive phase transition, while for γ = 2, Eq. 3 indicates an ideal relaxor behavior so-called complete diffuse phase transition (DPT) [43]. The plot of ln $(1/\varepsilon - 1/\varepsilon_m)$ versus ln(T - $T_m$) at 1 kHz of $Ba_{(1-x)}La_xTi_{0.89}Sn_{0.11}O_3$ ceramics is presented in the inset of each panel in Fig.6(a)-(d). The linear fitting of the experimental data based on Eq. 3, the slope of each plot determines the value of γ. The obtained values of γ for $Ba_{(1-x)}La_xTi_{0.89}Sn_{0.11}O_3$ samples were found to be 1.54, 1.89, 1.94, and 1.92 for BTSn, BLTSn 1.5, BLTSn 2.5 andBLTSn 3.5, respectively (as shown in the Fig.6(e)). These results indicate that the phase transition process in La-doped samples has typical relaxation dispersion features [44, 45]. The BLTSn 2.5 sample shows a maximum value of γ which is 1.94, indicating that adding La can improve the relaxation behavior of the BTSn sample. The enhanced relaxation feature is due to a structural disorder created by nanodomains (PNRs) as seen in the Raman spectra where the BLTSn 2.5 sample showed the highest FWHM [46].

To explain the diffusive behavior of the phase transition, the parameter $\Delta T_m$ is used to reveal the degree of diffuseness and could be determined by using the following equation:

$$\Delta T_m = T_{dev} - T_m \tag{4}$$

Where $T_{dev}$ is the temperature at which the dielectric constant begins to diverge from the Curie-Weiss law, and $T_m$ is the temperature corresponding to the maximum of the dielectric constant. The values of $\Delta T_m$ are found to be 22.1, 113.6, 125.4 and 176.9°C for BTSn, BLTSn1.5, BLTSn2.5 and BLTSn3.5, respectively. Among these results, the BLTSn3.5 sample displayed the highest value of $\Delta T_m$.

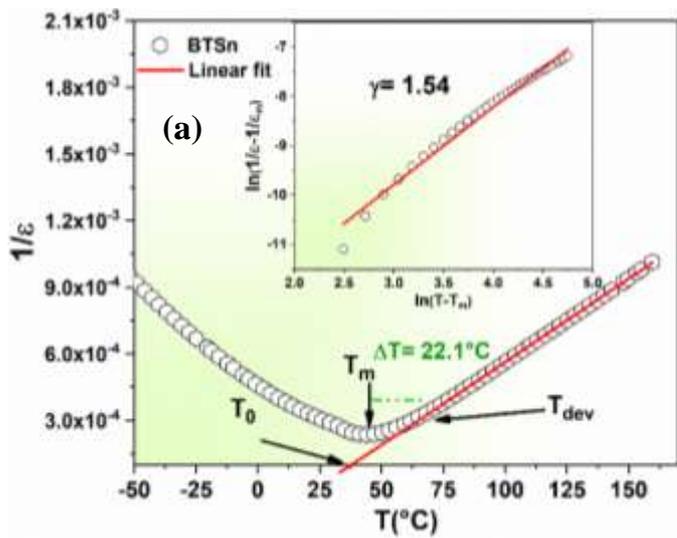
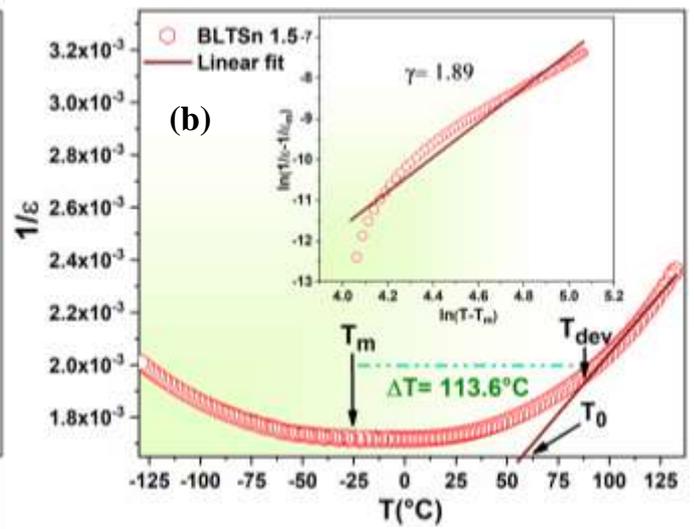
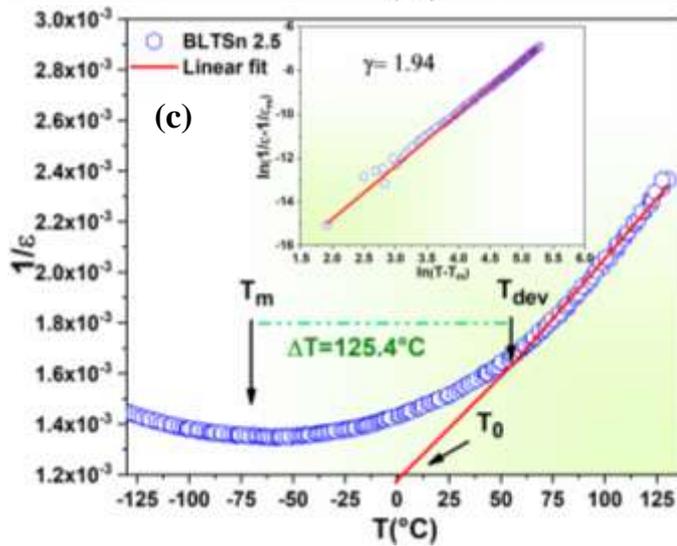
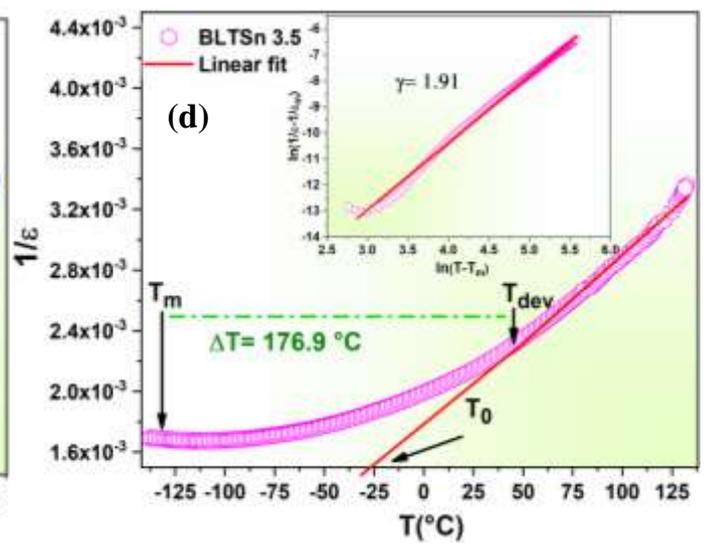
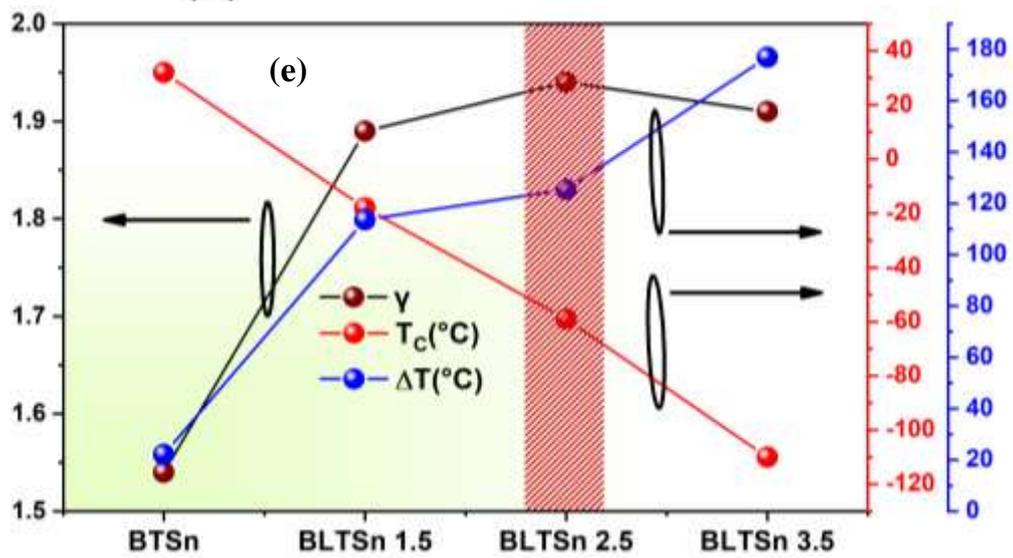

Figure 6 : (a-d) Temperature dependence of the $1/\varepsilon$ at 1 kHz (inset: the relationships between $\ln(1/\varepsilon - 1/\varepsilon_m)$ and $\ln(T - T_m)$ ) for all investigated ceramics. (e) Evolution of $\gamma$ $T_C$ and $\Delta T$ as a function of La concentration.

*3.2 Ferroelectric and energy storage properties*

Fig.7a-d illustrate the ferroelectric polarization versus electric field (P-E) loops of all investigated compositions BTSn, BLTSn1.5, BLTSn2.5, and BLTSn3.5 at 200 Hz under different applied electric fields, measured at room temperature. Each examined sample has been found to display typical ferroelectric P-E hysteresis loops. Meanwhile, to help visualize the curve changes more efficiently, the P-E loops for different compositions at room temperature are shown in Fig.7e. These figures show saturation curves in the P-E hysteresis loops for pure BTSn samples with relatively high remanent polarization, which provides high energy storage losses and low energy storage efficiency. On the other hand, the P-E hysteresis loops of La-doped samples are still not saturated, which indicates that La-doped ceramics could exploit for higher applied electric fields. Furthermore, Fig.7f shows the evolution of ferroelectric parameters ($P_r$, $P_{max}$, and $E_c$) with the concentration of La-doping. One can see that the ferroelectric coefficients are strongly dependent on the La-doping concentration. The P-E hysteresis loops become slimmer as La content increases, suggesting the transition from a normal ferroelectric to a relaxor type. This might be explained by the presence of polar nanoregions, which result in a thin P-E loop [47]. The findings show that La doping could enhance the ferroelectric performance of pure BTSn ceramics, and the XRD and dielectric analysis support this performance. In addition, the dielectric breakdown strength $E_b$ of La-doped samples was improved due to the sharply reduced grain size (according to the formula $E_b \propto G^{-1/2}$), as shown in the SEM micrographs [48].

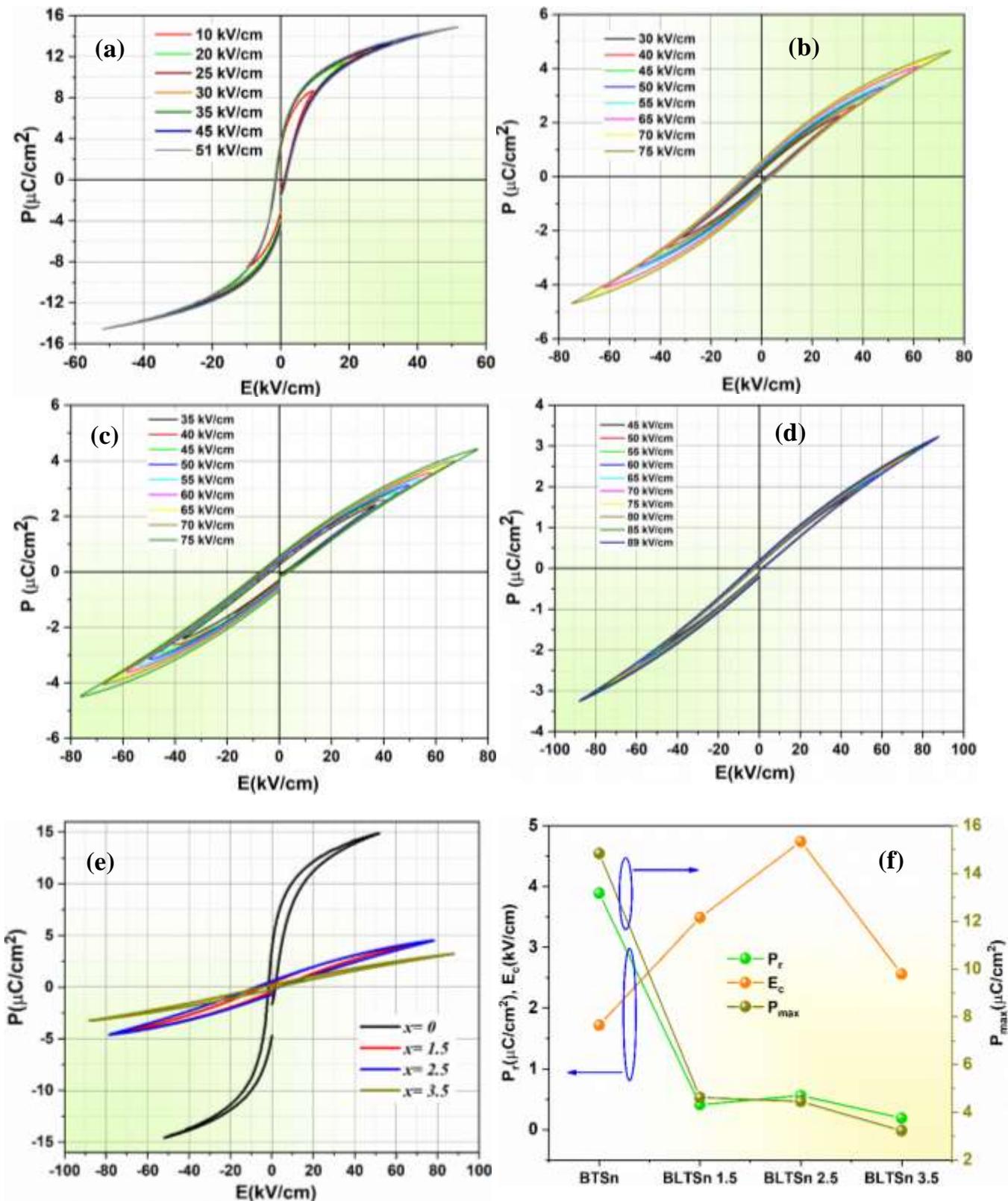

Figure 7: P-E hysteresis loops at different electric fields of (a) x=0, (b) x= 1.5, (c) x= 2.5 and (d) x= 3.5. (e) P-E hysteresis loops at RT of different samples. (f) $P_r$, $E_c$ and $P_{max}$ values of all samples.

To investigate the electrical energy storage performances of all studied samples, the P-E hysteresis loops were recorded at room temperature under different applied electric fields, as depicted in Fig.8. As known, the energy storage capacities can be easily obtained by integrating the area between the polarization axis and P–E curves discharge [49]. The different energy storage densities (total and recoverable) and energy storage efficiency could be calculated using the following equations [50]:

$$W_{tot} = \int_{0}^{Pmax} EdP, \tag{5}$$

$$W_{rec} = \int_{Pr}^{Pmax} EdP, \tag{6}$$

$$\eta(\%) = \frac{W_{rec}}{W_{tot}} * 100 = \frac{W_{rec}}{W_{rec}+W_{loss}} * 100, \tag{7}$$

Fig.8 shows the energy storage performances of $Ba_{(1-x)}La_xTi_{0.89}Sn_{0.11}O_3$ ceramics determined at room temperature. The extracted maximum recoverable energy density of $Ba_{(1-x)}La_xTi_{0.89}Sn_{0.11}O_3$ samples was found to be 142.8, 125.6, 158.8, and 118 mJ/cm$^3$ for BTSn, BLTSn1.5, BLTSn2.5 and BLTSn3.5, respectively. For the pure BTSn sample, the determined $W_{rec}$ was higher than the energy density reported in the literature for the same composition [51, 52]. The maximum recoverable energy density of 158.8 mJ/cm$^3$ was obtained in the BLTSn2.5 sample under an electric field of 75 kV/cm, higher than the undoped sample. The enhanced energy storage density could be due to the uniform and smaller grain size distribution observed in the SEM micrographs. Furthermore, the BLTSn2.5 exhibits the highest lattice strain, which could enhance energy storage performances by reorientating defect dipoles and suppressing the variation of polarization in the material [53, 54]. As indicated above, the La-doped samples could still support a high applied electric field, which leads to high energy storage properties.

In addition to a higher $W_{rec}$ value, improved energy storage efficiency is also necessary in practical applications. From Fig.8, the BLTSn3.5 shows the highest $\eta$, which results in a slimmer hysteresis loop due to small remnant polarization, leading to a lower energy loss density and therefore a high storage efficiency. The obtained energy storage efficiency of $Ba_{(1-x)}La_xTi_{0.89}Sn_{0.11}O_3$ ceramics was found to be 69.01, 80.85, 82.73 and 86.67% for BTSn, BLTSn1.5, BLTSn2.5 and BLTSn3.5, respectively. Notably, in the case of dielectric solid solutions, an opposite constraint relationship between $W_{rec}$ and $\eta$ has been noted. In contrast, the BLTSn2.5 sample exhibits a high efficiency of 82.73 %, along with the highest recoverable energy density. The combination of these results makes the Pb-free BLTSn2.5 ferroelectric sample a promising candidate for energy storage capacitor applications at room temperature.

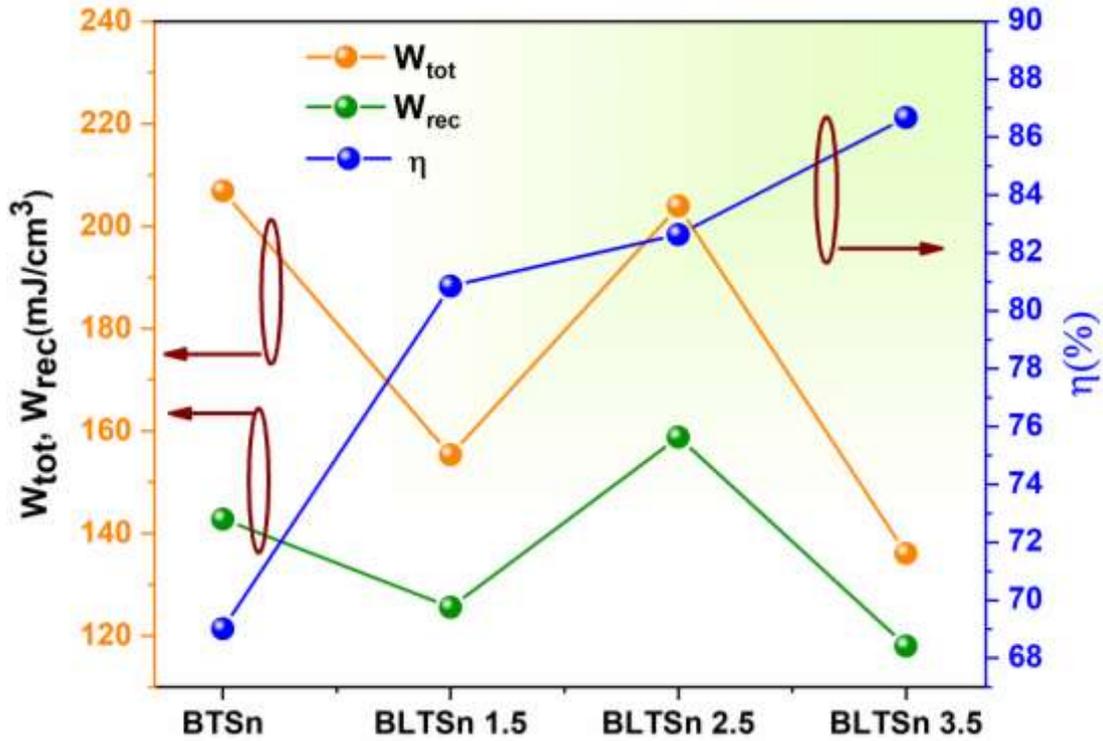

Figure 8: The relevant energy storage parameters of $Ba_{(1-x)}La_xTi_{0.89}Sn_{0.11}O_3$ ceramics at room temperature.

To set our findings in the literature, Table 1 compares the energy storage efficiency and recoverable energy density of several lead-free ceramics. Remarkably, stronger polarization is induced by increasing the applied electric field, which enhances energy storage capabilities. At almost the same electric field, our result (energy storage efficiency) surpasses the previously published results from numerous other lead-free ceramics. It is important to note that the elaboration methods, chemical composition, grain size engineering, and applied electric fields could affect the energy storage performances of dielectric materials.

Table 1: Comparison of energy storage performances of BLTSn ceramics with other lead-free ferroelectric ceramics.

| Samples | $W_{rec}$ (mJ/cm$^3$) | E (kV/cm) | $\eta$ (%) | Refs. |
|---|---|---|---|---|
| **BLTSn 2.5** | **158.8** | **75** | **82.73** | **This work** |
| **BLTSn 3.5** | **118** | **85** | **86.67** | **This work** |
| $Ba_{1-z}La_zTi_{0.95}Sn_{0.05}O_3$ | 210 | 40 | 30 | [19] |
| $Ba_{0.997}La_{0.003}TiO_3$ | 282 | 65 | 33.5 | [55] |
| $BaTi_{0.89}Sn_{0.11}O_3$ | 72.4 | 25 | 85.07 | [51] |
| $Ba_{0.85}Ca_{0.15}Zr_{0.10}Ti_{0.90}O_3$ | 62 | 25 | 72.9 | [56] |
| $[(Bi_{0.5}Na_{0.5})_{0.93}Ba_{0.07}]_{1-x}La_xTiO_3$ | 910 | 100 | 69 | [20] |
| $0.7Bi_{0.5}Na_{0.5}TiO_3-0.3SrTiO_3-xLa$ | 2005 | 170 | 82 | [57] |
| $Bi_{0.5}(Na_{0.84}K_{0.16})_{0.5}TiO_3-SrTiO_3$ | 120 | 60 | 48 | [58] |
| $0.92Ba_{0.85}Ca_{0.15}Zr_{0.1}Ti_{0.9}O_3-0.08Bi(Mg_{2/3}Ta_{1/3})O_3$ | 3400 | 440 | 93.87 | [59] |

Fig. 9a shows the P-E loops of the BLTSn2.5 ceramics tested at room temperature and in a frequency range of 200–1000 Hz under an electric field of 60 kV/cm. The P-E loops showed slight frequency variation, suggesting that the composition of BLTSn1.5 for energy storage has high-frequency stability. Fig. 9b depicts the variation of energy storage properties ($W_{rec}$ and $\eta$) of the BLTSn2.5 sample with the frequency. One can see that BLTSn2.5 ceramic exhibits excellent stability of $W_{rec}$ (125.6- 124.1 mJ/cm$^3$) and $\eta$ (80.85- 77.06 %) over the entire frequency range of 200-1000 Hz.

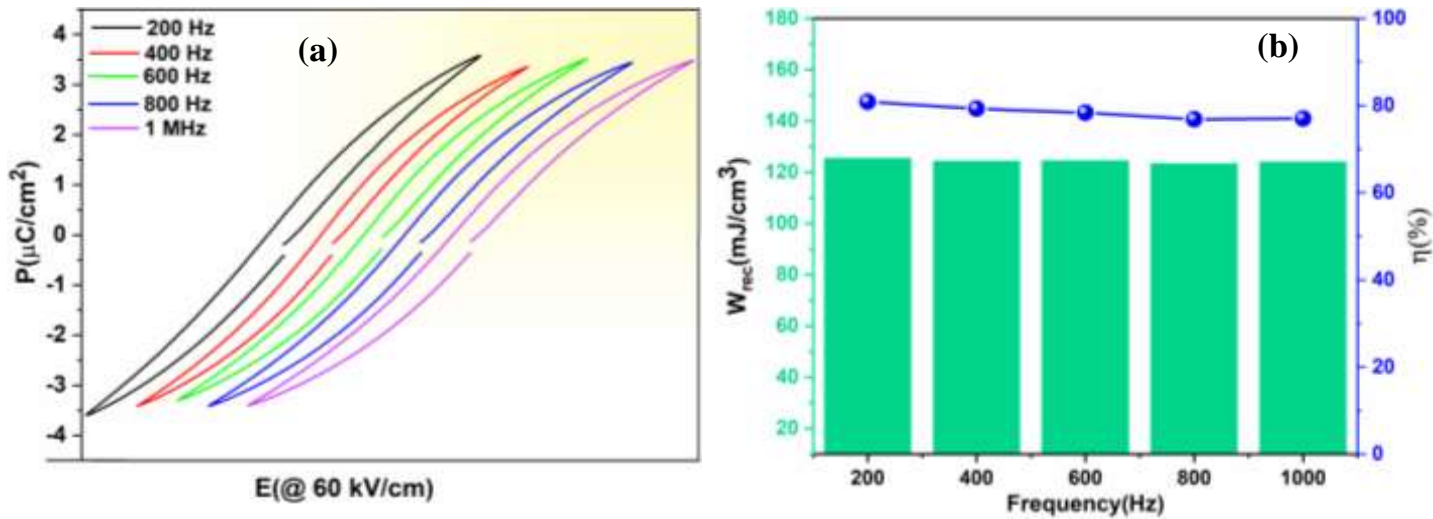

Figure 9: (a) P-E hysteresis loops and (b) W$_{rec}$ and $\eta$ of BLTSn 2.5 sample at different frequencies.

**Conclusion**

Lead-free Ba$_{(1-x)}$La$_x$Ti$_{0.89}$Sn$_{0.11}$O$_3$ ceramics were synthesized via a solid-state route and their structural, dielectric, ferroelectric, and energy storage properties have been investigated. After doping, all samples exhibit a pure perovskite structure, and no second phase is generated. The doped samples exhibit typical relaxor features, including broadened frequency dispersion and dispersed phase transition. The investigation of energy storage properties of the BLTSn2.5 sample demonstrated the enhanced W$_{rec}$ of 158.8 mJ/cm$^3$ along with a high $\eta$ of 82.73 % at room temperature. Besides, the sample exhibits excellent frequency stability in the 200-1000 Hz frequency range. We conclude that the BLTSn2.5 sample is a promising ceramic for environmentally friendly energy storage applications at room temperature.

**Acknowledgements**

This work was supported by the H-GREEN project (No 101130520) and the European Union Horizon 2020 Research and Innovation Action MSCA-RISE-MELON (No. 872631). The Slovenian Research and Innovation Agency (research core funding P2-0105 and research project N2-0212) is acknowledged.

**References**

[1] Q. Yuan, J. Cui, Y. Wang, R. Ma, et H. Wang, « Significant enhancement in breakdown strength and energy density of the BaTiO3/BaTiO3@SiO2 layered ceramics with strong interface blocking effect », *Journal of the European Ceramic Society*, vol. 37, nº 15, p. 4645-4652, dec. 2017, doi: 10.1016/j.jeurceramsoc.2017.06.028.

[2] S. Song *et al.*, « Ultrahigh electric breakdown strength, excellent dielectric energy storage density, and improved electrocaloric effect in Pb-free (1-x)Ba(Zr0.15Ti0.85)O3-xNaNbO3 ceramics », *Ceramics International*, vol. 48, nº 8, p. 10789-10802, apr. 2022, doi: 10.1016/j.ceramint.2021.12.295.

[3] Q. Li, F.-Z. Yao, Y. Liu, G. Zhang, H. Wang, et Q. Wang, « High-Temperature Dielectric Materials for Electrical Energy Storage », 2018.

[4] Y. Pérez-Martín *et al.*, « Electrocaloric effect, pyroelectric response and energy storage performance of lanthanum-modified PZT relaxor ferroelectric ceramic », *Physica B: Condensed Matter*, vol. 671, p. 415420, dec. 2023, doi: 10.1016/j.physb.2023.415420.

[5] X. Wang, « Glass additive in barium titanate ceramics and its influence on electrical breakdown strength in relation with energy storage properties », *Journal of the European Ceramic Society*, 2012.

[6] Y.-S. Zhang *et al.*, « Energy storage and charge-discharge performance of B-site doped NBT-based lead-free ceramics », *Journal of Alloys and Compounds*, vol. 911, p. 165074, aug 2022, doi: 10.1016/j.jallcom.2022.165074.

[7] X. Dong *et al.*, « Realizing enhanced energy storage and hardness performances in 0.90NaNbO3−0.10Bi(Zn0.5Sn0.5)O3 ceramics », *J Adv Ceram*, vol. 11, nº 5, p. 729-741, may 2022, doi: 10.1007/s40145-022-0566-6.

[8] Q. Li *et al.*, « Enhanced energy-storage performance of (1-x)(0.72Bi0.5Na0.5TiO3-0.28Bi0.2Sr0.7□0.1TiO3)-xLa ceramics », *Journal of Alloys and Compounds*, vol. 775, p. 116-123, feb. 2019, doi: 10.1016/j.jallcom.2018.10.092.

[9] Z. Yang, Y. Yuan, L. Cao, E. Li, et S. Zhang, « Relaxor ferroelectric (Na0.5Bi0.5)0.4Sr0.6TiO3-based ceramics for energy storage application », *Ceramics International*, vol. 46, nº 8, p. 11282-11289, jun 2020, doi: 10.1016/j.ceramint.2020.01.154.

[10] Z. Jiang, H. Yang, L. Cao, Z. Yang, Y. Yuan, et E. Li, « Enhanced breakdown strength and energy storage density of lead-free Bi0.5Na0.5TiO3-based ceramic by reducing the oxygen vacancy concentration », *Chemical Engineering Journal*, vol. 414, p. 128921, jun 2021, doi: 10.1016/j.cej.2021.128921.

[11] Z. Li *et al.*, « Largely enhanced energy storage performance of sandwich-structured polymer nanocomposites with synergistic inorganic nanowires », *Ceramics International*, vol. 45, nº 7, p. 8216-8221, may 2019, doi: 10.1016/j.ceramint.2019.01.124.

[12] W.-B. Li, D. Zhou, et L.-X. Pang, « Enhanced energy storage density by inducing defect dipoles in lead free relaxor ferroelectric BaTiO3-based ceramics », *Applied Physics Letters*, vol. 110, nº 13, p. 132902, mars 2017, doi: 10.1063/1.4979467.

[13] Y. Yu *et al.*, « High-temperature energy storage performances in (1-x)(Na0.50Bi0.50TiO3)-xBaZrO3 lead-free relaxor ceramics », *Ceramics International*, vol. 46, nº 18, p. 28652-28658, dec. 2020, doi: 10.1016/j.ceramint.2020.08.024.

[14] J. Gao *et al.*, « Enhancing dielectric permittivity for energy-storage devices through tricritical phenomenon », *Sci Rep*, vol. 7, nº 1, p. 40916, januar. 2017, doi: 10.1038/srep40916.

[15] Y. Yao *et al.*, « Large piezoelectricity and dielectric permittivity in BaTiO$_3$-xBaSnO$_3$ system: The role of phase coexisting », *EPL*, vol. 98, nº 2, p. 27008, apr. 2012, doi: 10.1209/0295-5075/98/27008.

[16] K. Banerjee et S. Asthana, « Resolution of ambiguity between the depolarization and ferroelectric–relaxor transition temperature through dielectric studies in lead-free perovskite K0.5Bi0.5TiO3 », *Materials Chemistry and Physics*, vol. 231, p. 344-350, jun 2019, doi: 10.1016/j.matchemphys.2019.04.043.

[17] F. D. Morrison, D. C. Sinclair, et A. R. West, « Doping mechanisms and electrical properties of La-doped BaTiO3 ceramics », *International Journal of Inorganic Materials*, vol. 3, nº 8, p. 1205-1210, dec. 2001, doi: 10.1016/S1466-6049(01)00128-3.

[18] R. Kumar, K. Asokan, S. Patnaik, et B. Birajdar, « Evolution of microstructure and relaxor ferroelectric properties in (LazBa1−z)(Ti0.80Sn0.20)O3 », *Journal of Alloys and Compounds*, vol. 687, p. 197-203, dec. 2016, doi: 10.1016/j.jallcom.2016.06.049.

[19]   R. Kumar, I. Singh, R. Meena, K. Asokan, B. Birajdar, et S. Patnaik, « Effect of La-doping on dielectric properties and energy storage density of lead-free Ba(Ti 0.95 Sn 0.05 )O 3 ceramics », *Materials Research Bulletin*, vol. 123, p. 110694, mars 2020, doi: 10.1016/j.materresbull.2019.110694.

[20]   J. Xu *et al.*, « Enhanced electrical energy storage properties in La-doped (Bi0.5Na0.5)0.93Ba0.07TiO3 lead-free ceramics by addition of La2O3 and La(NO3)3 », *J Mater Sci*, vol. 52, nº 17, p. 10062-10072, sept. 2017, doi: 10.1007/s10853-017-1209-0.

[21]   J. Yin *et al.*, « High energy-storage properties of Sr0.7Bi0.2-La TiO3 lead-free relaxor ferroelectric ceramics for pulsed power capacitors », *Ceramics International*, vol. 50, nº 20, p. 38562-38568, oct. 2024, doi: 10.1016/j.ceramint.2024.07.224.

[22]   J. Yin *et al.*, « Structurally Regulated Design Strategy of Bi0.5Na0.5TiO3-based Ceramics for », *ACS Applied Materials & Interfaces*, vol. 16, nº 25, p. 32367-32374, jun 2024, doi: https://doi.org/10.1021/acsami.4c04013.

[23]   Q. Huang, F. Si, et B. Tang, « The effect of rare-earth oxides on the energy storage performances in BaTiO3 based ceramics », *Ceramics International*, vol. 48, nº 12, p. 17359-17368, jun 2022, doi: 10.1016/j.ceramint.2022.02.299.

[24]   R. Muhammad, Y. Iqbal, et I. M. Reaney, « $BaTiO_3$ –Bi($Mg_{2/3}$ $Nb_{1/3}$ )$O_3$ Ceramics for High-Temperature Capacitor Applications », *J. Am. Ceram. Soc.*, vol. 99, nº 6, p. 2089-2095, jun 2016, doi: 10.1111/jace.14212.

[25]   S. Liu, L. Zhang, J. Wang, Y. Zhao, et X. Wang, « Structure, dielectric, ferroelectric and diffuse phase transition properties of the Ce, Ca hybrid doped BaTiO 3 ceramics », *Ceramics International*, vol. 43, p. S36-S42, augus 2017, doi: 10.1016/j.ceramint.2017.05.203.

[26]   R. L. Nayak, S. S. Dash, Y. Zhang, et M. P. K. Sahoo, « Enhanced dielectric, thermal stability, and energy storage properties in compositionally engineered lead-free ceramics at morphotropic phase boundary », *Ceramics International*, vol. 47, nº 12, p. 17220-17233, janur 2021, doi: 10.1016/j.ceramint.2021.03.033.

[27]   S. Chihaoui, L. Seveyrat, V. Perrin, I. Kallel, L. Lebrun, et H. Khemakhem, « Structural evolution and electrical characteristics of Sn-doped Ba0.8Sr0.2TiO3 ceramics », *Ceramics International*, vol. 43, nº 1, p. 427-432, januar. 2017, doi: 10.1016/j.ceramint.2016.09.176.

[28]    R. Kumar, K. Asokan, S. Patnaik, et B. Birajdar, « Evolution of microstructure and relaxor ferroelectric properties in (La$_z$Ba$_{1-z}$)(Ti$_{0.80}$Sn$_{0.20}$)O$_3$ », *Journal of Alloys and Compounds*, vol. 687, p. 197-203, dec. 2016, doi: 10.1016/j.jallcom.2016.06.049.

[29]    M. Deluca *et al.*, « High-field dielectric properties and Raman spectroscopic investigation of the ferroelectric-to-relaxor crossover in BaSn$_x$Ti$_{1-x}$O$_3$ ceramics », *Journal of Applied Physics*, vol. 111, nº 8, p. 084102, apr. 2012, doi: 10.1063/1.3703672.

[30]    R. Kumar, K. Asokan, S. Patnaik, et B. Birajdar, « Combined effect of oxygen annealing and La-doping in broadening the phase transition of Ba(Zr$_{0.2}$Ti$_{0.8}$)O$_3$ ceramics », *Journal of Alloys and Compounds*, vol. 737, p. 561-567, mars 2018, doi: 10.1016/j.jallcom.2017.11.344.

[31]    A. Feteira, D. C. Sinclair, et J. Kreisel, « Average and Local Structure of $(1-x)$BaTiO$_3$ $-x$ LaYO$_3$ ($0 \leq x \leq 0.50$) Ceramics », *Journal of the American Ceramic Society*, vol. 93, nº 12, p. 4174-4181, dec. 2010, doi: 10.1111/j.1551-2916.2010.04006.x.

[32]    A. Sati *et al.*, « Effect of structural disorder on the electronic and phononic properties of Hf doped BaTiO3 », *J Mater Sci: Mater Electron*, vol. 30, nº 10, p. 9498-9506, may 2019, doi: 10.1007/s10854-019-01281-5.

[33]    I. A. Souza *et al.*, « Structural and dielectric properties of Ba$_{0.5}$Sr$_{0.5}$(Sn$_x$Ti$_{1-x}$)O$_3$ ceramics obtained by the soft chemical method », *Journal of Alloys and Compounds*, vol. 477, nº 1-2, p. 877-882, may 2009, doi: 10.1016/j.jallcom.2008.11.042.

[34]    S. Singh, P. Singh, O. Parkash, et D. Kumar, « Structural and relaxor behavior of (Ba$_{1-x}$La$_x$)(Ti$_{0.85}$Sn$_{0.15}$)O$_3$ ceramics obtained by solid state reaction », *Journal of Alloys and Compounds*, vol. 493, nº 1-2, p. 522-528, mars 2010, doi: 10.1016/j.jallcom.2009.12.148.

[35]    J. Gao, W. Liu, L. Zhang, F. Kong, Y. Zhao, et S. Li, « Enhanced dielectric and ferroelectric properties of the hybrid-doped BaTiO3 ceramics by the semi-solution method », *Ceramics International*, vol. 47, nº 11, p. 15661-15667, jun 2021, doi: 10.1016/j.ceramint.2021.02.136.

[36]    H. Yang *et al.*, « Novel BaTiO$_3$-Based, Ag/Pd-Compatible Lead-Free Relaxors with Superior Energy Storage Performance », *ACS Appl. Mater. Interfaces*, vol. 12, nº 39, p. 43942-43949, sept. 2020, doi: 10.1021/acsami.0c13057.

[37]   S. Y. Chen *et al.*, « Enhanced energy storage properties of Ba0.85Ca0.15Zr0.1Ti0.9O3 ceramics doped with BiFeO3 », *Journal of Alloys and Compounds*, vol. 936, p. 168015, mars 2023, doi: 10.1016/j.jallcom.2022.168015.

[38]   G. Liu *et al.*, « Dielectric, ferroelectric and energy storage properties of lead-free (1-x)Ba0.9Sr0.1TiO3-xBi(Zn0.5Zr0.5)O3 ferroelectric ceramics sintered at lower temperature », *Ceramics International*, vol. 45, nº 12, p. 15556-15565, augus 2019, doi: 10.1016/j.ceramint.2019.05.061.

[39]   H. Sun *et al.*, « Large energy storage density in BiFeO3-BaTiO3-AgNbO3 lead-free relaxor ceramics », *Journal of the European Ceramic Society*, vol. 40, nº 8, p. 2929-2935, july. 2020, doi: 10.1016/j.jeurceramsoc.2020.03.012.

[40]   Z. Chen *et al.*, « Novel BCZT-based ceramics with ultrahigh energy storage efficiency and outstanding high temperature fatigue endurance and stability for practical application », *Ceramics International*, vol. 49, nº 22, p. 34520-34528, nov. 2023, doi: 10.1016/j.ceramint.2023.08.087.

[41]   P. Bharathi et K. B. R. Varma, « Grain and the concomitant ferroelectric domain size dependent physical properties of Ba$_{0.85}$Ca$_{0.15}$Zr$_{0.1}$Ti$_{0.9}$O$_3$ ceramics fabricated using powders derived from oxalate precursor route », *Journal of Applied Physics*, vol. 116, nº 16, p. 164107, oct. 2014, doi: 10.1063/1.4900494.

[42]   X. Ren *et al.*, « Regulation of energy density and efficiency in transparent ceramics by grain refinement », *Chemical Engineering Journal*, vol. 390, p. 124566, jun 2020, doi: 10.1016/j.cej.2020.124566.

[43]   J. Zhang, J. Zhai, et Y. Zhang, « Microwave and infrared dielectric response in BaTiO$_3$ based relaxor ferroelectrics », *Physica Status Solidi (a)*, vol. 208, nº 12, p. 2853-2860, dec. 2011, doi: 10.1002/pssa.201127272.

[44]   A. Zhang *et al.*, « Significant improvement in energy storage for BT ceramics via NBT composition regulation », *Journal of Alloys and Compounds*, vol. 968, p. 172255, dec. 2023, doi: 10.1016/j.jallcom.2023.172255.

[45]   Y. Wang *et al.*, « A-site compositional modulation in barium titanate based relaxor ceramics to achieve simultaneously high energy density and efficiency », *Journal of the

*European Ceramic Society*, vol. 41, nᵒ 13, p. 6474-6481, oct. 2021, doi: 10.1016/j.jeurceramsoc.2021.05.052.

[46] Q. Yuan *et al.*, « Simultaneously achieved temperature-insensitive high energy density and efficiency in domain engineered BaTiO3-Bi(Mg0.5Zr0.5)O3 lead-free relaxor ferroelectrics », *Nano Energy*, vol. 52, p. 203-210, oct. 2018, doi: 10.1016/j.nanoen.2018.07.055.

[47] W. Kleemann, « The relaxor enigma — charge disorder and random fields in ferroelectrics », *J Mater Sci*, vol. 41, nᵒ 1, p. 129-136, janua. 2006, doi: 10.1007/s10853-005-5954-0.

[48] D. Zeng *et al.*, « Achieving high energy storage density in BaTiO3-(Bi0.5Li0.5)(Ti0.5Sn0.5)O3 lead-free relaxor ferroelectric ceramics », *Journal of Alloys and Compounds*, vol. 937, p. 168455, mars 2023, doi: 10.1016/j.jallcom.2022.168455.

[49] S. Khardazi *et al.*, « Enhanced thermal stability of dielectric and energy storage properties in 0.4BCZT-0.6BTSn lead-free ceramics elaborated by sol-gel method », *Journal of Physics and Chemistry of Solids*, vol. 177, p. 111302, jun 2023, doi: 10.1016/j.jpcs.2023.111302.

[50] Q. Li *et al.*, « Enhanced energy-storage properties of (1-x)(0.7Bi0.5Na0.5TiO3-0.3Bi0.2Sr0.7TiO3)-xNaNbO3 lead-free ceramics », *Ceramics International*, vol. 44, nᵒ 3, p. 2782-2788, febr. 2018, doi: 10.1016/j.ceramint.2017.11.018.

[51] S. Merselmiz *et al.*, « High energy storage efficiency and large electrocaloric effect in lead-free BaTi0.89Sn0.11O3 ceramic », *Ceramics International*, vol. 46, nᵒ 15, p. 23867-23876, oct. 2020, doi: 10.1016/j.ceramint.2020.06.163.

[52] M. Zahid *et al.*, « Enhanced near-ambient temperature energy storage and electrocaloric effect in the lead-free BaTi0.89Sn0.11O3 ceramic synthesized by sol–gel method », *J Mater Sci: Mater Electron*, vol. 33, nᵒ 16, p. 12900-12911, jun 2022, doi: 10.1007/s10854-022-08233-6.

[53] X. Gong *et al.*, « Enhancing energy storage efficiency in lead-free dielectric ceramics through relaxor and lattice strain engineering », *Journal of Materiomics*, vol. 10, nᵒ 6, p. 1196-1205, nov. 2024, doi: 10.1016/j.jmat.2023.12.006.

[54] S. Sun *et al.*, « Large piezoelectricity and potentially activated polarization reorientation around relaxor MPB in complex perovskite », *Journal of the European Ceramic Society*, vol. 42, nº 1, p. 112-118, janua. 2022, doi: 10.1016/j.jeurceramsoc.2021.09.065.

[55] V. S. Puli, P. Li, S. Adireddy, et D. B. Chrisey, « Crystal structure, dielectric, ferroelectric and energy storage properties of La-doped BaTiO3 semiconducting ceramics », *J. Adv. Dielect.*, vol. 05, nº 03, p. 1550027, sept. 2015, doi: 10.1142/S2010135X15500277.

[56] S. Merselmiz *et al.*, « Thermal-stability of the enhanced piezoelectric, energy storage and electrocaloric properties of a lead-free BCZT ceramic », *RSC Adv.*, vol. 11, nº 16, p. 9459-9468, 2021, doi: 10.1039/D0RA09707A.

[57] Z. Xu, Y. Chen, et N. Cheng, « Improved energy storage properties and stability of La modified 0.7Bi0.5Na0.5TiO3-0.3SrTiO3 ceramics », *Ceramics International*, vol. 50, nº 14, p. 25948-25954, july. 2024, doi: 10.1016/j.ceramint.2024.04.337.

[58] K. A. Aly, V. Athikesavan, E. Ranjith Kumar, et M. M. Ebrahium, « Preparation and study of La-doped bismuth sodium potassium titanate -strontium titanate piezoelectric ceramics to enhance energy storage properties », *Ceramics International*, vol. 50, nº 7, p. 11676-11687, apr. 2024, doi: 10.1016/j.ceramint.2024.01.071.

[59] D. Han *et al.*, « Superior energy storage properties of (1-x)Ba0.85Ca0.15Zr0.1Ti0.9O3-xBi(Mg2/3Ta1/3)O3 lead-free ceramics », *Journal of Alloys and Compounds*, vol. 946, p. 169300, jun 2023, doi: 10.1016/j.jallcom.2023.169300.